\documentclass[fleqn,twoside]{article}

\usepackage{amsmath,amssymb}

\usepackage{gc}
\usepackage{psfig,rotate}

\def\v#1,{{\bf#1},}
\def\j#1{{\it#1\/}}

\heads{Popov et al} 
      {Young close isolated compact objects} 
      
\begin{document}
\twocolumn[
\bigskip
      
\Title{YOUNG CLOSE ISOLATED COMPACT OBJECTS}

\Author{S.B. Popov\foom 1 and
         M.E. Prokhorov\foom 2}%
       {Sternberg Astronomical Institute,
              Universitetski pr.\,13, 119899 Moscow.}%

\Author{M. Colpi\foom 3}%
   {Universit\'a degli Studi di Milano Bicocca,
	Piazza della Scienza 3, I-20126 Milano.}
        
\Author{A. Treves\foom 4}%
   {Universit\'a degli Studi dell'Insubria,
	via Vallegio 11, 22100 Como.}
        
\Author{R. Turolla\foom 5}%
    {Universit\'a degli Studi di Padova, 
	via Marzolo 8, 35131 Padova.}

\Abstract{
We suggest that the seven radio-quiet isolated neutron stars observed with ROSAT 
are young cooling objects associated  to recent
near-by supernova explosions which formed runaway stars and  
the Local Bubble, affecting the topology of the interstellar medium 
in the vicinity of the Sun (within a few hundred parsecs).
In the aftermath of these explosions, 
a few black holes might have been formed, 
according to the local initial mass function.
We thus  discuss the possibility of determining  approximate positions
of close-by isolated black holes using  data on runaway stars
and simple calculations of binary evolution and disruption.}


]
\email 1 {polar@sai.msu.ru}
\email 2 {mike@sai.msu.ru}
\email 3 {Monica.Colpi@mib.infn.it}
\email 4 {treves@mi.infn.it}
\email 5 {turolla@pd.infn.it}

X-ray observations of neutron stars and black holes of stellar origin 
have been probing, over the entire Galaxy,  properties of their interior and  
of their environment. 
But the recent discovery of a number of isolated neutron stars in the outskirts of the
Sun
provides a new tool to study how they evolve and from
where they originate.

In this paper at first we discuss the properties of the seven ROSAT radio-quiet
isolated neutron star (INS) candidates showing that the cooling hypothesis for
their X-ray emission is the most viable
(see [19] and [36] for recent reviews 
and [40]  for data on the latest candidate). 
Then we try to connect these seven young INSs with different objects
in the solar vicinity. 
Finally we discuss a possibility to observe close-by isolated black
holes (IBHs).
 

\section{Neutron stars}

It is now widely believed that young INS can appear, in the Galactic disk, as sources of different
nature: radio pulsars, soft-gamma repeaters (SGRs), anomalous X-ray pulsars
(AXPs), radio-quiet compact X-ray sources in supernova (SNa) remnants.
%

Here we will focus on seven ROSAT INSs, for which it was suggested that they
are close-by sources showing no sign of pulsar activity.

\subsection{Accretion versus cooling}

In the early 70's it was suggested by Shvartsman [29], [30] 
and Ostriker et al.  [21]
that INSs and IBHs can be observable
due to accretion of the interstellar medium (ISM).
INSs can also be  observed when they are young and incandescent due to thermal emission
of their hot surface mainly in soft X-rays.

Soon after the discovery of the  first of the seven sources [38], it was clear that 
two different  mechanisms  could equally explain
the nature of the emission: accretion onto  the (old) INS
[35] or cooling of the young INS (see [19], [20] for a recent discussion).
Early in the study of the detectability of INSs, it seemed plausible that
{\it accretors} by far outnumber {\it coolers},
since the stage during which a young cooling INS is hot enough to be
observed in
soft X-rays is very short, about 1 Myr or less, against few Gyr lifetime of a
typical accreting NS. However, later it was recognized  that accreting INSs
might be rare objects and this holds if NSs acquire at birth very high peculiar
velocities 
$\sim 200$ km s$^{-1}$ (as observed in the pulsar sample [5], [17]). 
This  would imply an exceedingly low
accretion rate, and in turn a very low intrinsic luminosity, 
undetectable with present capabilities. 
Population synthesis studies [24]
have show that with the inclusion of a velocity distribution function that accounts for the
observed high kick velocities, most
of the INSs are now in the stage of {\it ejector} 
(see [15] for
detailed explanation of NS evolutionary stages or [4] for
a brief description). 
It is worth noting that 
if  strongly magnetized INSs with very high spatial
velocities are situated in low density ISM regions
then it is possible that most of them spend significant part
of their lives as {\it georotators}, because the light cylinder radius,
$R_l=c/\omega$, is larger than the radius of gravitational capture, 
$R_G=2GM/v^2$. 
Such INSs leave the {\it ejector} stage earlier 
than those that live in a normal ISM [34]. 

The periods $P$ of the observed objects 
(for four of the ROSAT INSs)  have been found to fall  in a rather narrow interval 
of 5-23 seconds,  and they  
could be explained in both hypothesis. In the accretion scenario 
one has to allow for magnetic field decay
[14], [39], because at a constant field $B\sim 10^{12}$ G, the periods of an old {\it accretor}
is much longer [16], [26]. 
Under the cooling hypothesis there are no restrictions on the period  $P$, though its value
can provide
clues on the magneto-rotational evolution of the objects.
Indeed, in the case of young cooling objects 
periods as long as such observed can be explained if the INSs are 
 {\it magnetars} [6], i.e. has
very strong magnetic field. 
The large braking implied by a high field would lead to an "accelerated" 
slow-down of the pulsar.
But to explain the clustering observed,  
the magnetic field later has to decay 
(as customarily advocated in magnetars for explaining the 
high quiescent luminosity of SGRs and AXPs [33]).
This  would lead to a saturation
of $P$ occurring when the decay is advanced enough to almost freeze the action of
braking torques [3].
If ROSAT INSs are magnetars or are their close relatives (or descendant) 
than the fraction of magnetars among all NSs can be
higher than  previously estimated.
Is the cooling hypothesis, which seem  preferable on theoretical ground,  
testable from observation? Clues would come from any  direct determination of the
peculiar velocity for the seven sources. Accretion can be certainly excluded
when the velocity exceeds 40-60 km s$^{-1}$ under the most favorable conditions in the ISM.

With the determination of the proper motion of one of the seven sources, 
RX J1856-3754,
(that implies a velocity at birth of 200 km s$^{-1}$ [37]) 
it is now rather clear that the cooling hypothesis holds for this sources, 
and it is tempting to consider the possibility  that the emission of  all
the seven ROSAT sources have the same nature. 
This calls again for a statistical study. 
To this purpose, in [25] we compared both hypothesis ({\it accretors } and {\it coolers}) 
in an attempt to
explain the observed Log N -- Log S distribution for the ROSAT INSs.
We obtained that at low fluxes ($< 10^{-13}$ erg cm$^{-2}$ s$^{-1}$)
{\it accretors} outnumber {\it coolers}. 
But at brighter fluxes, where the seven ROSAT
INSs are situated, {\it coolers}  can be more abundant. 
Interestingly, we found that to explain these seven objects in terms of cooling 
INSs it is necessary to assume that the 
spatial density of NSs
is about half an order of magnitude higher than what inferred  from radio pulsars
statistics.  
Are we then living in an over-dense region or is it a global feature?
If global,  we have to assume that most 
young NSs do not pass through the stage of a radio pulsar, or that this
stage is shorter than we have estimated. Such hypothesis finds support in recent
observations
[12], but the fraction of radio-quiet INSs 
in the whole galactic population of NSs is a number quite uncertain.
Most probably it is not as high as it is necessary to interpret 
the "magnificent seven" 
ROSAT INSs.
So it is more probable that such an over-density is a local phenomenon, 
both in space ($<1$ kpc) and in time ($<100$ Myr), 
and we next we will explore in more details the neighborhoods of the Sun.


\subsection{The ROSAT sources and the Gould Belt}  
 
 In connection with young
INSs and IBHs we are mostly interested in massive stars
and recent star formation activity close to the Sun (in our future
calculations
of {\it coolers}  we plan to include the realistic distribution of 
young stellar complexes around the Sun).
In that sense the solar vicinity is dominated by the Gould Belt and
related OB-associations
(see [23] for a detailed description of the Gould Belt).

 The Gould Belt is a disk-like structure. Its inclination to the Galactic
plane is about 18$^\circ$; the diameter is about 750-1000 pc, and its center is
located about 150-250 pc from the Sun. 

 The age of the Gould Belt is estimated to be about 30-70 Myr. 
This implies that the most massive stars are now about 7-10 $M_{\odot}$
and that recently there was a period of frequent 
explosions of massive stars, producing NSs and BHs, with nearly constant
rate [18]. The SNa rate in the Gould Belt is about
20-30 per Myr [13].

 We see direct consequences of these explosions in the form of
runaway stars.
In 700 pc around the Sun 56 runaway stars are known [10]. 
Only few of them can result from star-star interactions.
Others are products of SNa explosions in binary systems.
For some of them, the corresponding compact objects have been identified
or suggested [11], [13], [37].

 Clearly the seven ROSAT INSs can be the outcome  of 
recent SNae in the Gould Belt
and related OB-associations. In that case we can easily explain the
increased
rate of NS formation around the Sun,
and it is not necessary to assume very a high fraction of radio-silent
young INSs overall in the Galaxy.
Also one can predict the existence of more radio-quiet INSs within 
 1~kpc around the Sun, which are remnants of $> 50$ recent SNae. For objects 
with age $\sim 1$ Myr, 
the corresponding X-ray dim sources can be discovered (see for example 
objects HIP 22061, 29678 in table 5 in [10]), but as long as the NSs receive 
large kick velocities at birth it is very difficult to predict their 
present positions. For BHs the situation can be opposite as their kicks 
could be much lower.

All these explosions should leave their imprints on the structure of
the local ISM. Indeed, several local cavities are observed.
The most well known is the Local Bubble [28].
It was suggested [31], [18] that the Local
Bubble is a result of 3-6 (or even more) recent SNa explosions.  
We argue that  at least some of the 
"magnificent seven" 
sources are remnants of these recent explosions.


\section{Black holes and the Gould Belt}

 SNa explosions produce not only NSs, but also BHs.
Having dozens of SNae in the solar vicinity during the last 10 Myr 
we can expect several BHs to have formed during the same period in the
solar
neighborhood. 

 Usually it is accepted that BHs are one order of magnitude less abundant
than NSs. This estimate comes from the critical mass for BH formation.
If this mass is about 35~$M_{\odot}$ then the fraction of BHs is about 10\%
(see discussion on BH fraction in [7], [8]).
So we can expect about 5 BHs correlated to 56 runaway stars.
If there are 20-30 SNae per Myr in the Gould Belt, than we can expect
$\sim 6$-$12$ IBHs younger than few Myr.
Kick velocity for BHs is unknown, but it is reasonable to assume, that it is
much lower than for NSs. 
If it is so, all these BHs still should be around us.

 IBHs are not expected to be bright objects.
Close-by IBH can be observed due to accretion from the ISM ([30], [9]),
or due to a micro-lensing effect ([1], see also [22]).
That is why it is important to know their positions on the sky.
Close massive runaway stars give us a chance to calculate an
approximate positions of close-by young IBHs.
Among runaway stars we can distinguish the most massive:
$\lambda$ Cep, $\zeta$ Pup, HIP 38518 and  $\xi$ Per [10]. 
Their masses are larger than $\sim 33$ $M_{\odot}$. It means, that the
companion (actually the {\it primary} in the original binary) was even more
massive
on the main sequence stage. So, the most likely product 
of the explosion of such a massive star  should be a BH.

If the present velocities 
of runaway stars are known, one can estimate
their ages and places of birth. This has been done by Hoogefwerf et al.
[10].
To calculate the present position of a BH we have to know the binary
parameters, i.e., the 
masses of stars before the explosion, the BH mass, the  eccentricity
of the
orbit before the explosion, the 
orbit orientation, and finally the kick velocity of the BH. 
Some parameters can be inferred from the  observation of the
secondary star. Also
we can give a zero kick to the BH and
zero orbital eccentricity. Below we briefly comment
on that choice. Other parameters should be varied within assumed ranges.

As we do not know the exact mechanism of SNa explosion
any value of a BH kick velocity
would be speculative. But in all mechanisms BH kicks
should be smaller than those of NSs. 
In particular, in [27] the authors argue
that the magneto-rotational mechanism of SNae explosion
(suggested in 70's by Bisnovatyi-Kogan) is the most
favored from the point of view of mass distribution of compact objects.
In this mechanism fastly rotating protoNS form increasing 
superstrong toroidal magnetic
field (up to $10^{17}$ G), which drives the envelope ejection.
In this case BHs should receive kicks much smaller than those of NSs.
We assume BH kick to be zero (but note, that in [8] it was suggested that 
in disrupted binaries one has to expect mostly low mass BHs, $M_{bh}\sim 3\,
M_{\odot}$, which receive a modest kick $\sim 50$ km s$^{-1}$ at their birth).
In  the binary systems which are the progenitor of runaway stars we can
expect circularization of orbits due to tidal interaction, 
i.e. eccentricity in such binaries should be zero.
As far as the present velocities of secondaries 
are high we can expect close systems ($a\sim 1000\, R_{\odot}$)
with nearly ideal circularization.

Given these parameters and still leaving undetermined the orientation of
the orbital plane of a binary, what remains to discuss are the masses of
the primary component before the explosion and of the BH. In the case of
two massive companions and in the most probable situation, the mass of the
primary before the explosion is close to that of the secondary.  These
circularized systems with equal mass stars will survive the explosion if
the kick is strictly zero, since less than one half of the total mass will
be swept out. So, for systems producing run-away stars the most probable
primary's mass is as close as  that requested for binary
disruption, i.e.:

\begin{equation}
M_1=M_2+2\cdot M_{bh}.
\end{equation} 
That leads to restrictions on the orbital separation: it should be
larger than about 100 $R_{\odot}$ to avoid mass transfer. On the other
hand it has to be smaller
than $\sim $2000 $R_{\odot}$, to guarantee that the velocity of the
secondary
component after SNa explosion is in agreement with observations.

Strong mass loss due to stellar winds also
leads to relatively low masses for BH progenitor (before the explosion),
as far as more massive stars loose
mass due to stellar winds faster than low mass stars,
so for two stars with initially very
different masses this difference will become smaller before the explosion.

Results of [7], [8]
suggest that massive stars with $M>40M_{\odot}$ produce BHs without
SNae, so for such cases binaries will survive BH formation. This
argument once again brings us to require the lowest masses for
the  primaries that are permitted 
according to the condition of binary disruption. 
Masses should then fall within a relatively narrow interval
close to critical mass of BH formation.

 Masses of BHs are determined now for nearly a dozen of candidates
in binary systems, mainly with low-mass companions. 
Most of these determinations
are concentrated around $\sim 7$-$10$ $M_{\odot}$, 
but the potentially cover a relatively wide range
from 3 up to 50 solar masses [2], [27] (see the theoretical
expectations of the BH mass spectrum, which is different from the observed
one due to selection effects [8]). 

Given these constrains we are in the process of studying the dynamics of
the runaway BHs in the solar proximity and work is in progress.

  
\section{Conclusions}

 We conclude that the seven radio-quiet ROSAT INSs can be connected with
recent SNa explosions, which produced nearby runaway stars and
peculiar features in the local ISM including the Local Bubble.

 We suggest a way to find approximate positions of close IBHs from
knowledge of nearby 
runaway stars and from calculations of binary disruptions. We estimate
a number of close IBHs as $>5$ with ages $<3$-$4$ Myr.

\Acknow
{ We thank Luca Zampieri for discussions.
S.P. thanks Universit\'a degli Studi dell'Insubria, Universit\'a degli
Studi
di Padova and Universit\'a degli Studi di Milano Bicocca for hospitality
and support.

The work of S.P. and M.P. 
was supported by RFBR (grants 01-02-06265, 01-15(02)-99310).}


\small
\begin{thebibliography}{99}
\bibitem{1}D.P. Bennet et al., 
``Gravitational microlensing events due to stellar mass black holes'';
astro-ph/0109467.
\bibitem{2}A.M. Cherepashchuk, \j{Phys. Usp.} \v39, 753 (1996).
\bibitem{3}M. Colpi, U. Geppert and D. Page,
\j{ApJ} \v529, L29 (2000).
\bibitem{4}
M. Colpi, A. Possenti, S. Popov and F. Pizzolato,
\j{in:} ``Physics of Neutron Star Interiors'', 
eds. D. Blaschke, N.K. Glendenning and A. Sedrakian,
Springer--Verlag, Berlin, p.441, 2001; astro-ph/0012394. 
\bibitem{5}J.M. Cordes and D.F. Chernoff,
\j{ApJ} \v505, 315 (1998).
\bibitem{6}R.C. Duncan and C. Tompson, \j{ApJ} \v392, L9 (1992). 
\bibitem{7}C.L. Fryer, \j{ApJ} \v522, 413 (1999).
\bibitem{8}C.L. Fryer and V. Kalogera, \j{ApJ} \v544, 548 (2001).
\bibitem{9}Y. Fujita, I. Susumi, T. Nakamura, T. Manmoto and K.E. Nakamura,
\j{ApJ} \v495, L85 (1998).
\bibitem{10}R. Hoogerwerf, J.H.J. de Bruijne and P.T. de Zeeuw,
\j{A\&A}  \v365, 49 (2001).
\bibitem{11}R. Hoogerwerf, J.H.J. de Bruijne and P.T. de Zeeuw,
\j{ApJ} \v544, L133 (2001).
\bibitem{12}E.V. Gotthelf and G. Vasisht,
\j{in:} Proceedings of IAU Coll. 177,   
"Pulsar Astronomy -- 2000 and Beyond",
eds. M. Kramer, N. Wex and N. Wielebinski, 
\j{ASP Conf. Series}  \v202, 699 2000. 
\bibitem{13}I.A. Grenier, \j{A\&A} \v364, L93 (2000).
\bibitem{14}
D.Yu. Konenkov and S.B. Popov, 
\j{PAZh} \v23, 569 (1997); astro-ph/9707318.
\bibitem{15}
V.M. Lipunov, ``Astrophysics of Neutron Stars'',
Springer--Verlag, Berlin, 1992.
\bibitem{16}
V.M. Lipunov and S.B. Popov, \j{AZh} \v72, 711 (1995).
\bibitem{17}A.G. Lyne and D.R. Lorimer,
\j{Nature} \v369, 127 (1994).
\bibitem{18}J. Ma\'iz-Apell\'aniz, 
\j{ApJ} \v560, L83 (2001).
\bibitem{19}C. Motch, \j{in:} Proceedings of ``X-ray Astronomy '999 --- Stellar
Endpoints, AGN and the Diffuse Background'', 
eds. G. Malaguti, G. Palumbo and N. White, Gordon \& Breach (Singapore), 2001
; astro-ph/0008485.
\bibitem{20}R. Neuh\"auser and J.E. Tr\"umper,
\j{A\&A} \v343, 151 (1999).
\bibitem{21}J.P. Ostriker, M.J. Rees, and J. Silk, 
\j{Astroph. Letters} \v6, 179 (1970).
\bibitem{22}B. Paczynski, 
``Can HST measure the mass of the isolated neutron star RX J185635-3754?'';
astro-ph/0107443.
\bibitem{23}W. P\"oppel, 
\j{Fund. Cosm. Phys.} \v18, 1 (1997).
\bibitem{24}
S.B. Popov, M. Colpi, 
A. Treves, R. Turolla, V.M. Lipunov and M.E. Prokhorov,
\j{ApJ} \v530, 896 (2000).
\bibitem{25}
S.B. Popov, M. Colpi, M.E. Prokhorov,
A. Treves and R. Turolla,
\j{ApJ} \v544, L53  (2000).
\bibitem{26}M.E. Prokhorov, S.B. Popov and A.V. Khoperskov,
``The period distribution of old accreting isolated neutron stars'',
to appear in \j{A\&A}; astro-ph/0108503.
\bibitem{27}M.E. Prokhorov, K.A. Postnov,
``Why NS and BH mass distribution is bimodal?'',
to appear in \j{Odessa Astr. Publ.}; astro-ph/0110176.
\bibitem{28}D.M. Sfeir, R. Lallement, F. Crifo and B.Y. Welsh,
\j{A\&A} \v346, 785 (1999).
\bibitem{29}V.F. Shvartsman,
\j{AZh} \v47, 824 (1970).
\bibitem{30}V.F. Shvartsman,
\j{AZh} \v48, 479 (1971).
\bibitem{31}R.K. Smith and D.P. Cox,
\j{ApJ Supp.} \v134, 283 (2001).
\bibitem{32}T.M. Tauris et al.,
\j{ApJ} \v428, L53 (1994).
\bibitem{33}C. Thompson and R.C. Duncan,  \j{ApJ}, \v473, 322 (1996)
\bibitem{34}O.D. Toropina, M.M. Romanova, Yu.M. Toropin and R.V.E. Lovelace,
 ``Propagation of Magnetized Neutron Stars Through the Interstellar
Medium'', to appear in \j{ApJ}; astro-ph/0105422.
\bibitem{35}A. Treves and M. Colpi, A\&A \v241, 107 (1991).
\bibitem{36}
A. Treves,  R. Turolla, S. Zane and M. Colpi, 
\j{PASP} \v112, 297 (2000).
\bibitem{37}F.M. Walter,
\j{ApJ} \v549, 433 (2001).
\bibitem{38}F.M. Walter, S.J. Wolk and R. Neuh\"auser, 
\j{Nature} \v379, 233 (1996).
\bibitem{39}J. Wang, \j{ApJ} \v486, L119 (1997).
\bibitem{40}L. Zampieri  et al.,
\j{A\&A} \v378, L5 (2001).
\end{thebibliography}
\end{document}